\documentstyle[prl,preprint,aps]{revtex}
\begin{document}
\draft
\title{
Quantum conductivity corrections in two dimensional long-range
disordered systems
with strong spin-orbit splitting of electron spectrum.}
\author{A.P.~Dmitriev, I.V.~Gornyi, and V.Yu.~Kachorovskii}
\address{A.F.~Ioffe Physical-Technical Institute,\\
26 Polytechnicheskaya str., Saint Petersburg, 194021, Russia.\\
e-mail: dmitriev@vip1.ioffe.rssi.ru}
\maketitle
\tightenlines

\begin{abstract}
We study quantum corrections to conductivity in a 2D system with a
smooth random potential and strong spin-orbit splitting of
the spectrum. We show that the interference correction
is positive and down to the very low temperature can exceed the negative
correction related to electron-electron interaction.
We discuss this result in the context of the problem of
the metal-insulator transition in Si-MOSFET structures.

\end{abstract}

%\narrowtext
\vspace{1cm}

Recent experiments on the high-mobility Si-MOSFET \cite{krav&} and
other \cite{GaAs} structures
have demonstrated the possibility of the metal-insulator transition (MIT)
in two-dimensional (2D) electronic systems. It has been found
that when electron density $n$ becomes larger than
some critical value $n_c$ the system exhibits a metallic
behavior. This result is rather surprising because it
contradicts the common belief,
based on the predictions of the one-parameter scaling theory
\cite{AALR}, that all
electronic states in a 2D disordered system should
be localized.

The theory \cite{AALR}  deals with  noninteracting electrons.
Meanwhile, the peculiarity of the observed MIT is that it
occurs at a very low
electron density when the Coulomb interaction energy is larger than the Fermi
energy.
This hints that the role of the electron-electron interaction is
very important
and the  transition
is determined by the interplay between the interaction and disorder.
First, the arguments supporting the possibility of
the interaction
driven MIT in a 2D system were put
forward in the old  work  by Finkelstein \cite{fink}, where the
 renormalization
of the  triplet  channel
coupling constant played a key role. The experiments
\cite{krav&,GaAs}
stimulated  intensive discussion
of Finkelstein approach
anew \cite{cast}.

The calculations  \cite{fink,cast} were
performed under the assumption that the spin-orbit interaction
is absent in the system.
However, according to the work \cite{spinsplitt},
used in the experiments Si-MOSFET
structures (in these structures MIT is most conspicuous) have a strong
spin-orbit splitting of the electron energy spectrum
due to the built-in electric field of an asymmetric quantum well.
Moreover, the value of the spectrum splitting $\Delta$
is estimated in \cite{spinsplitt,pud} to be larger than
$\hbar/\tau$, where $\tau$ is the elastic
mean free time. In this situation the
contribution of the triplet channel is
strongly suppressed and the
 Finkelstein approach may turn out not relevant to the concrete
 experimental situation.
This circumstance has  initiated  theoretical investigation
of quantum conductivity corrections in systems with an arbitrary relation
between the splitting $\Delta$ and
$\hbar/\tau$ \cite{sk,lg}.
The calculations have been performed
for the case of a short-range impurity potential. It has been shown
that
at a rather small splitting
the interference correction
to the conductivity
changes sign and becomes antilocalizing.
For a strong splitting
$\Delta \tau \gg \hbar$  the correction
arises due to the interference of two electronic waves with
the total spin equal to zero and  is given by
\begin{equation}
\delta\sigma_{wl}=
\frac{e^2}{4 \pi^2 \hbar}
\ln{\frac{\tau_\phi}{\tau}}.
\label{eq1}
\end{equation}
Here $\tau_\phi $ is the phase breaking time, which at
sufficiently low temperatures $T$ is governed solely by electron-electron
collisions, and is inversely proportional to  $T$ \cite{AA}.
The result (\ref{eq1}) is analogous to the well-known result
\cite{Hikami&} for the case of a strong spin-orbit scattering
on a random potential.

Generally speaking,  the change in a sign of the quantum correction
to the conductivity indicates the possibility of MIT .
However, it is commonly believed (see for example recent discussion
\cite{AMDC,Belitz,castell}) that such a change in
the interference correction due to spin-orbit interaction does not
provide a metallic behavior of a 2D system.
Indeed, besides the positive interference correction there
exists also a
negative Aronov-Altshuler quantum correction related to electron-electron
exchange interaction
in the singlet channel, which is not strongly affected by
spin-orbit interaction
\cite{AA}
\begin{equation}
\delta\sigma_{ee}
=-\frac{e^2}{2 \pi^2 \hbar}
\ln{\frac{\hbar}{T\tau}}.
\label{eq2}
\end{equation}
The Hartree contribution to conductivity correction
is absent in this situation, because
it arises due to the
interaction  in the
triplet channel and is suppressed in the presence of a strong
spin-orbit coupling \cite{AA1}.

At low temperatures the absolute value of $\delta\sigma_{ee}$
exceeds the quantity in Eq. (\ref{eq1}). Therefore the
total conductivity correction is negative and logarithmically diverges
as $T$ decreases, which corresponds to insulating behavior.

In the presence of the valley degeneracy Eq.(\ref{eq1}) is valid at
$\tau_\phi \gg \tau_v$, where $\tau_v$ is the characteristic time of
intervalley scattering. In the opposite limiting case $\tau_\phi \ll
\tau_v$ valleys are independent,
interference corrections from each valley  are added together,
and the expression (\ref{eq1}) is multiplied
by the number of valleys $N_v$. At the same time the expression for
$\delta\sigma_{ee}$ does not contain
$N_v$ (Refs. \cite{AA,Fuk}) and the temperature
dependence of the total correction
is given by
 \begin{equation}
\delta\sigma_{tot}
=\frac{e^2}{2 \pi^2 \hbar}
\left(\frac{N_v}{2}-1\right)
\ln{\frac{\hbar}{T\tau}}.
\label{eq3}
\end{equation}
Hence it follows that the existence of several valleys
diminishes the relative role of the localization effects and may lead to
the antilocalization in the temperature interval, where $\tau_\phi \ll
\tau_v$.  However, at $N_v=2$ (which corresponds to the case of Si-MOSFET
structures \cite{krav&}) the effect of the valley degeneracy still does not
lead to a metallic behavior.
Indeed, as follows from (\ref{eq3}),  the total
conductivity correction  does not depend on the temperature in this case.
This result is correct if we completely neglect the transitions
between valleys,
i.e. to the zeroth order in the parameter $\tau_\phi/\tau_v$. One can show,
however, that already to the  first order in this parameter the
conductivity slowly decreases  with decreasing temperature, i.e. taking
into account of arbitrary weak intervalley transitions leads to insulating
behavior.

In this  work we investigate the effect of the strong
spin-orbit splitting of the spectrum ($\Delta\tau\gg\hbar$) on
the quantum  corrections to the conductivity of electrons moving in a
{\it smooth} random potential. We do not intend to explain
the metal-insulator transition. All  we are going to do is
to calculate the  quantum
conductivity corrections in the range of large conductance values,
i.e., far from the transition ($n\gg n_c$).
The inequality $\Delta \tau \gg \hbar$  allows us to
neglect the contributions of the
triplet channels in the interference correction as well as
in the interaction correction.
The results  differ essentially from
the case of a {\it short-range} impurity potential:
  when the spectrum splitting is strong enough the positive
interference conductivity correction
is two times larger than the value given by Eq. (\ref{eq1})
down to very low temperatures.
As a consequence, the total conductivity correction
(with the electron-electron interaction
and the valley degeneracy taken into account)
turns out to be {\it antilocalizing}
and the system exhibits a metallic behavior:
the resistivity decreases with temperature decreasing.

This result is related to the fact that, in contrast to the case
of point-like impurities, in a  smooth
potential transitions between spin subbands are strongly suppressed
and each valley breaks up into two independent subsystems.
Thus in the case of two valleys the quantity $N_v$ in the
expression (\ref{eq3}) effectively is equal to $4$ instead of $2$.
Note, that in experiments \cite{krav&} the disorder
was created mainly by ionized impurities located in the
oxide layer and therefore the random potential was long-ranged.

We restrict ourselves to a consideration of the
system at high densities $n$, in which the Fermi
energy $E_F$ is much greater than the Coulomb energy.
 We consider
the case of a single valley first. The term in the Hamiltonian of
noninteracting electrons, which is responsible for
the spin-orbit splitting of the energy spectrum (so-called
Rashba term) is given by
$\alpha([\hat\sigma\times{\bf k}]\cdot {\bf n})$.
Here the quantity $\alpha$ is a constant
characterizing the strength of the spin-orbit
interaction, the operator $\hat\sigma$ is a vector, consisting of
the Pauli matrices, $\hbar {\bf k}$ is the momentum of an electron
 and the vector ${\bf n}$ is the unit vector normal to
the 2D layer. The energy spectrum and wavefunctions of an electron
with the effective mass $m$ read

\begin{equation}
E^{\pm}({\bf k})=\frac{\hbar^2 k^2}{2m} \pm \alpha k \:,\; \;
\phi^{\pm}_{\bf k}({\bf r})
=\exp(i {\bf kr}) \chi^{\pm}_{\bf k}, \; \;
\chi^{\pm}_{\bf k}=\frac{1}{\sqrt{2}}
\left(
\begin{array}{c}
1 \\
\pm i e^{i \varphi_{\bf k}}
\end{array}
\right )
\label{spektr}
\end{equation}

Spinors $\chi^{\pm}_{\bf k}$ depend on the polar
angle $\varphi_{\bf k}$ of the wave vector ${\bf k}$,
and describe two states with the spins polarized parallel to the
vectors $\pm [{\bf k}\times {\bf n}]$, respectively.
Thus the system is divided into two subsystems (branches),
"$+$" and "$-$", each one having the spin of an
electron  rigidly connected to the
momentum.

A random smooth impurity potential $U({\bf r})$ has
a correlation function $K(r)=\langle U({\bf r})U(0) \rangle $,
falling off at the scale $d\gg k_F^{-1}$.
The potential is assumed to be weak enough, so that the following
inequalities are held: $E_F\tau\gg \hbar$ and $\tau \gg d/v_F$.
The presence of this potential leads to transitions
both within each branch, and between different branches.
The respective (intrabranch and interbranch) times of
these transitions are given by the following expression
\begin{equation}
\frac{1}{\tau_{\mu \nu}}=\frac{2\pi}{\hbar}
\int \frac{d^2 {\bf q}}{(2\pi)^2} K_{\mu \nu}({\bf q})
\delta[E^{\mu}({\bf k}) - E^{\nu}({\bf k}-{\bf q})],
\label{tau*}
\end{equation}
where $K_{\mu \nu}({\bf q})= | \langle \chi^{\mu}_{\bf k} |
\chi^{\nu}_{{\bf k}-{\bf q}} \rangle |^2 K(q) $, indices
$\mu$ and $\nu$ signify the type of the branch
($+$ or $-$), and the function $K(q)$ is the Fourier-transform
of the potential correlator  with
a characteristic scale $d^{-1}$. Further we assume that
the spin-orbit splitting  $\Delta = 2 \alpha k_F$
is less than the Fermi energy $E_F$. In this case we
can set $\tau_{++}=\tau_{--}=\tau$,  $\tau_{+-}=\tau_{-+}=\tau_{*}$.
The minimal transferred
momentum needed for the transition between branches is determined
by the delta-function in (\ref{tau*})
and  is equal to $2 m \alpha/\hbar $. If in addition to the
inequality $\Delta \tau \gg \hbar$ the stronger one
\begin{equation}
m\alpha d \gg \hbar^2
\label{uslovie}
\end{equation}
is fulfilled, then interbranch transitions are
suppressed compared to transitions
within one branch. In particular, for the case of the
potential created by ionized impurities located
at distance $d$ from the 2D layer, the correlation function
$K(q)\sim \exp(-2qd)$, and  the interbranch
transition time is given by
\begin{equation}
\tau_*=4 (k_F d)^2
\exp\left(\frac{4 m \alpha d}{\hbar^2}\right)\,\tau .
\label{tauexp}
\end{equation}
The factor $(k_F d)^{2}$ in this expression appears
due to the orthogonality of the spinors corresponding to
different branches and to identically directed momenta.
It is worth noting that when an electron is scattered by
a point-like impurity potential, an  arbitrary
momentum transfer is possible ($K(q)= const $) and as a consequence,
the value of $\tau_*$ is of the same order
as the intrabranch transition time.

Let us consider the interference conductivity correction,
neglecting for a while the electron-electron interactions.
On the time scale less than $\tau_*$ interbranch transitions
are absent and the system may be treated as consisting of
two independent subsystems, corresponding to two uncoupled
branches. Therefore for at $\tau_\phi \ll \tau_* $ the
interference conductivity correction is equal to  sum of
the interference corrections of two branches. In each branch
this correction is given by
$$
\delta\sigma^{+}_{wl}=\delta\sigma^{-}_{wl}=(1/2)({e^2}/{2 \pi^2 \hbar})
\ln(\tau_\phi/\tau_{tr}),
$$
where $\tau_{tr}\sim(k_F d)^2 \tau $
is the transport scattering time.
Calculations are similar to those in the works
\cite{bergman,bhatt} where the weak localization correction
has been derived in the case of a smooth potential
in the absence of spin-orbit splitting.
However, in our case the result is two times less
in the absolute value  and  has
the opposite sign, i.e., we have the {\it antilocalizing}
correction. The factor $1/2$ arises due to the fact that
calculating the contribution of one branch, we take into
account only one spin state. The difference in the sign of
the conductivity correction is caused by spinor transformation
properties: under  rotation by an angle of $2\pi$ the spin
wavefunction is multiplied by $-1$. Indeed, the correction
$\delta\sigma_{wl}$ is related to the effective change of
the backscattering amplitude due to the interference of two electron
waves, propagating along closed paths in the opposite
directions. In our case the spin is always perpendicular to
the momentum and is rotated by an angle of $\pi$ along
one of the interfering paths and by $-\pi$  along oncoming one.
As a result the relative rotation of spins of the two interfering waves
is equal to $2\pi$. This  leads to the change of the
correction sign \cite{my}. The total interference correction
is given by
\begin{equation}
\delta\sigma_{wl}= \delta\sigma^{+}_{wl}+\delta\sigma^{-}_{wl}
=\frac{e^2}{2 \pi^2 \hbar}
\ln\frac{\tau_\phi}{\tau_{tr}}.
\label{wl}
\end{equation}

When the relation between $\tau_{\phi}$ and $\tau_*$
is arbitrary calculations can be performed  using
the method of solution of the Cooperon equation in
multiband systems, proposed in \cite{Hikami&} and
further developed in Ref. \cite{nikita}.
The result reads
\begin{equation}
\delta\sigma_{wl}=\frac{e^2}{4 \pi^2 \hbar} \left[
\ln\frac{\tau_\phi }{\tau_{tr}} +
\ln\frac{\tau_\phi  \tau_* }{(2\tau_\phi + \tau_*) \tau_{tr}}\right] .
\label{totwl}
\end{equation}
At $\tau_\phi \ll \tau_* $ this expression turns into
Eq. (\ref{wl}). In the limit $\tau_{\phi} \gg \tau_*$,
when on the phase breaking time a large number of transitions between
branches takes place ,
the main contribution in Eq. (\ref{totwl})
arises from the first term. This contribution is analogous
to Eq. (\ref{eq1}) of the point-like potential case.

Eq. (\ref{totwl}) can be clarified with help of
the following transparent arguments. When the spin-orbit
splitting is large enough, two waves propagating along
the closed path in the opposite directions,
should belong to the same branch
 between
two succeeding scatterings (although, after scattering
both waves can change the type of the branch together).
Moreover, only those processes are important, in which
both waves at the beginning and at the end of
the path also belong to one branch.
In the opposite case two waves acquire different phases
and do not interfere.
Then the conductivity correction due to interference
of the two waves is proportional to the number of returns of
the particle to the initial point and initial branch.
Let us estimate this number in the diffusion
approximation. The corresponding equations have the form
\begin{eqnarray}
\frac{\partial c }{\partial t} - D \Delta c &=&
\delta({\bf r})\delta(t) + \frac {c^\prime - c}{\tau_*} - \frac
{c}{\tau_\phi} , \nonumber
\\
\frac{\partial c^\prime }{\partial t} - D \Delta c^\prime &=&
 \frac {c - c^\prime}{\tau_*} - \frac
{c^\prime}{\tau_\phi} .
\label{C}
\end{eqnarray}
Here $c({\bf r}, t) $ and $c^\prime({\bf r}, t) $ are
the probability densities of finding  a particle at the
moment $t$ in the point ${\bf r}$ in the initial and
the second branch respectively, $D=v_F^2\tau_{tr}/2$
is a diffusion coefficient. The Fourier-transform
of $c({\bf r}, t) $ is the sum of two
pole terms:
$$
c_{q,\omega}=\frac{1}{2}\left ( \frac{1}{Dq^2-i\omega+\tau_\phi^{-1}}
+\frac{1}{Dq^2-i\omega+\tau_\phi^{-1}+2 \tau_{*}^{-1}} \right ) \; .
$$
The number of returns to the initial branch, we are interested in,
is proportional to the integral
$\int^{\infty}_{\tau_{tr}} c({\bf r}=0,t) dt $,
calculating  which, one gets the sum of two logarithms
in Eq. (\ref{totwl}).

Now let us discuss the role of a weak electron-electron
interaction in our problem. As usual, it leads to
 Aronov-Altshuler quantum conductivity correction $\delta\sigma _{ee}$ and
determines the phase breaking time at low temperatures \cite{AA}.
In our case electron-electron collisions can also
cause transitions between spectrum branches.

It can be shown that the conductivity correction
$\delta\sigma_{ee}$ is still given by Eq. (\ref{eq2}).
This is due to the fact, that
$\delta\sigma_{ee}$ depends on  the total
probability of particle to propagate
from the initial point
to the final one, independently from the type of the branch
at the end of the path. This probability does not depend
on $\tau_*$ which can be demonstrated using Eqs. (\ref{C})
(in which terms containing $\tau_\phi$ should be omitted).
Indeed, the quantity $\tau_*$ goes away
from the
equation for the sum of $c$ and $c^\prime$.
In the other words, the fact of existing of
two conducting subsystems, on the one hand, doubles the
conductivity correction,
but on the other hand, the Coulomb interaction in this case
is screened two times stronger due to the same fact.
The insensitiveness of $\delta\sigma_{ee}$ to the existence of two branches
is analogous to the mentioned earlier
fact, that the $\delta\sigma_{ee}$ does not depend on the number of valleys
$N_v$.

As far  as the time of phase breaking due to
electron-electron interaction is concerned, this quantity is also not very
sensitive to
the spin-orbit interaction and one can use the
formula from Ref. \cite{AA}:
$$\frac {1}{\tau_\phi} \sim \frac{T}{E_F\tau_{tr}}
\ln{\frac{E_F \tau_{tr}}{\hbar}}.$$

Transitions between two branches due to electron-electron
collisions can be neglected, since
the characteristic time of such transitions, $\tau_*^{ee}$,
is much larger than  $\tau_\phi$ and $\tau_*$
at low temperatures. It is due to the fact, that the
minimal momentum transfer in the interbranch transition
is much greater than inverse mean free path and
when calculating $\tau_*^{ee}$ the
diffusion approximation fails. As a result $\tau_*^{ee}$
occurs to be inversely proportional to the squared
temperature  \cite{AA}, so it appears to be larger than
either $\tau_\phi$, and $\tau_*$.

Let us now introduce the valley degeneration, assuming
that $\tau_v \gg \tau_*, \tau_\phi$. In this case the total quantum
conductivity correction is a sum of Eq. (\ref{eq2})
(with $\tau$ replaced by $\tau_{tr}$) and Eq. (\ref{totwl}),
multiplied by $N_v$
\begin{equation}
\delta\sigma_{tot}=\frac{e^2}{2 \pi^2 \hbar} \left[
 \frac{N_v}{2}
\ln\frac{\tau_\phi^2  \tau_* }{(2\tau_\phi + \tau_*) \tau_{tr}^2 } -
\ln\frac{\hbar}{T\tau_{tr}}  \right] .
\label{final}
\end{equation}
At the temperatures
\begin{equation}
T>T_*\sim \frac{E_F}{\ln(E_F \tau_{tr}/\hbar)}
\frac{\tau_{tr}}{\tau_*},
\label{Tmin}
\end{equation}
the phase breaking time appears to be shorter than
interbranch transition time
($\tau_\phi < \tau_* $).
In this range of  temperatures
for the depending on $T$ part of the total correction
we get
\begin{equation}
\delta\sigma_{tot}=\frac{e^2}{2 \pi^2 \hbar}
(N_v-1) \ln{\frac{\hbar}{T\tau_{tr}}}.
\label{total}
\end{equation}
 From this expression one can see that for $N_v=2$
the conductivity increases with decreasing temperature,
i.e., a metallic behavior takes place.

In the region $\tau_\phi>\tau_*$ the
conductivity continues to  increase very slowly (for $ N_v = 2 $) and
reaches the value
\begin{equation}
\delta\sigma_{tot}=\frac{e^2}{2 \pi^2 \hbar}
 \ln\frac{E_F \tau_* }{\hbar} .
\label{final1}
\end{equation}
Only at very low temperatures, when the transitions between valleys come into
play ( $\tau_\phi \sim \tau_v$ ),
the conductivity begins to decrease
logarithmically.

We now present some estimations. According to \cite{pud}
in the real experimental situation
$\alpha\approx5\cdot10^{-6}K\,cm$, $m=0.2 m_e$,
and $d\sim10^{-5}cm$. Under these conditions the exponent
in Eq. (\ref{tauexp}) is large, so that $\tau_*\gg\tau_{tr}$.
Then at $n=(5 \div 10)\cdot10^{11}cm^{-2}$ for
$T_*$ we have values of the order of several $mK$, which
is less than the lowest temperature  ($20mK$),
used in the experiments \cite{krav&}.
Thus a metallic behavior of the 2D electron gas in Si-MOSFETs
at low temperatures and high densities $n \gg n_c$
could be possibly explained by proposed above mechanism.

Authors are grateful to V.M.Pudalov,
M.I.Dyakonov, N.S.Averkiev, and L.E.Golub for
useful discussions.
This work was supported by Russian Foundation for
Basic Research, Swedish Royal Academy of Sciences
and INTAS (grant 96-196).

\end{document}